\documentclass{aa}
\usepackage{times}
\usepackage{graphics}
\begin{document}

%   \thesaurus{09     % A&A Section 9: un
%              (02.03.3;   % Convection
%               02.12.1;   % Line: formation
%               02.18.7;   % Radiative transfer
%               06.01.1;   % Sun: abundances
%               06.07.2;   % Sun: granulation
%               06.16.2)}  % Sun: photosphere
%
   \title{Line formation in solar granulation:}

   \subtitle{V. Missing UV-opacity and the photospheric Be abundance}

   \author{Martin Asplund
          }

\titlerunning{Line formation in solar granulation: V. The photospheric Be abundance}
\authorrunning{Asplund}

   \institute{Research School of Astronomy and Astrophysics,
%Australian National University,
Mount Stromlo Observatory,
Cotter Road, Weston, ACT 2611, Australia.
\email{martin@mso.anu.edu.au}
             }

   \offprints{martin@mso.anu.edu.au}

%   \date{Received: yes; accepted: yes}
   \date{{\em Accepted for Astronomy \& Astrophysics}}

%%%%%%%%%%%%%%%%%%%%%%%%%%%%%%%%%%%%%%%%%%%%%%%%%%%%%%%%%%%%%%%%%%%%
   \abstract{
The possibility of unaccounted for opacity sources in
the UV for late-type stars has often been invoked to explain
discrepancies between predicted and observed flux distributions
and spectral line strengths. Such missing UV-opacity could
among other things have a significant impact on abundance determination
for elements whose only relevant spectral features are
accessible in this wavelength region, such as Be.
Here, the study by Balachandran \& Bell (1998) is re-visited in
the light of a realistic 3D hydrodynamical solar model atmosphere
and the recently significantly downward revised solar O abundance obtained
with the same model atmosphere.
The amount of missing UV-opacity, if any, is quantified by
enforcing that the OH A-X electronic lines around 313\,nm produce the same
O abundance as the other available diagnostics: OH vibration-rotation
and pure rotation lines in the IR, the forbidden [O\,{\sc i}] 630.0 and 636.3\,nm
lines and high-excitation, permitted O\,{\sc i} lines.
This additional opacity is then applied for the synthesis of the
Be\,{\sc ii} line at 313.0\,nm to derive a solar photospheric Be abundance
in excellent agreement with the meteoritic value, thus re-enforcing
the conclusions of Balachandran \& Bell.
The about 50\% extra opacity over accounted for opacity sources
can be well explained by recent calculations by
the Iron Project for photo-ionization of Fe\,{\sc i}.
   }

   \maketitle

%
%________________________________________________________________

\section{Introduction}

Among other things, the light elements lithium, beryllium and boron
offer an opportunity to extract information about stellar interiors
and evolution. Due to their fragile nature, these elements are
destroyed by nuclear processing when brought to sufficiently high
temperatures by stellar convection or other mixing events.
Since the destruction of Li, Be and B occurs at somewhat different
temperature regimes ($\sim 2.6 \cdot 10^6$\,K for $^7$Li,
$\sim 3.5 \cdot 10^6$\,K for Be and $\sim 5 \cdot 10^6$\,K for B),
the relative amount of depletion of these elements can function as
sensitive probes how deep the stellar mixing has proceeded.
Indeed there is extensive information available on light element
depletion in stars of different spectral types (e.g. Deliyannis et al. 2000),
confirming the presence of additional mixing over that
predicted by the standard mixing length theory for convection during
the pre-main sequence and main sequence evolution.
Among the invoked explanations for the observed depletion patterns,
slow rotationally-induced mixing has perhaps received the most attention
recently given the reasonable agreement with observations for its
predictions over a range of stellar environments (Deliyannis et al. 2000).

Neither has our Sun been immune to light element depletion.
The observed photospheric Li abundance is about a factor of 140 lower than
measured in CI-chondrite meteorites (M\"uller et al. 1975; Kiselman 1997;
Asplund et al. 2003a; Lodders 2003),
a difference which is far beyond the
uncertainties involved in the photospheric and meteoritic abundance determinations.
Until recently, it was believed that also Be and B were depleted in the solar
photosphere by factors of about 1.8 (Chiemlewski et al. 1975) and
1.9 (Kohl et al. 1977; Kiselman \& Carlsson 1996), respectively,
relative to the meteoritic evidence.
This standard picture was challenged by Balachandran \& Bell (1998,
see also Bell et al. 2001), who attempted to calibrate the
since-long suspected missing UV-opacity (e.g. Vernazza et al. 1976;
Gustafsson \& Bell 1979)
by enforcing that the same
oxygen abundance should be determined from the OH A-X electronic lines
around the 313\,nm region, where the crucial Be\,{\sc ii} lines are
located, as the more reliable IR vibration-rotation counterparts of OH
(log\,$\epsilon_{\rm O} = 8.91$\footnote{
On the customary logarithmic abundance scale defined
to have a hydrogen abundance of log$\,\epsilon_{\rm H}=12.00$}
in the case of the 1D Holweger-M\"uller solar model atmosphere).
Their investigation indicated a missing opacity for these
wavelengths of about 60\% over known opacity sources,
which they attributed to photo-ionization of Fe\,{\sc i}.
Indeed, subsequent analysis of detailed atomic calculations performed
within the framework of the Iron Project (Hummer et al. 1993) indicated that
this hypothesis was quite plausible (Bell et al. 2001).
When accounting for this missing continuous opacity in the
synthesis of the Be\,{\sc ii} doublet at 313\,nm, Balachandran \& Bell (1998)
found that the resulting
photospheric Be abundance was indistinguishable from the commonly adopted
meteoritic Be abundance of log\,$\epsilon_{\rm Be} = 1.41$
(Lodders 2003).
Furthermore, both the photospheric and meteoritic B abundances have recently
been revised, bringing them into apparent agreement within their (substantial)
uncertainties (Cunha \& Smith 1999; Zhai \& Shaw 1994).
Current thinking thus implies that while Li in the solar convection
zone has been significantly destroyed, neither Be nor B have been
depleted relative to their proto-solar values. As indicated above,
such a depletion pattern requires additional non-standard mixing,
although the precise nature of this mixing is still eluding astronomers.

It is important to realise that the influential analysis of
Balachandran \& Bell (1998) hinges on at least six crucial assumptions:
\begin{enumerate}
\item the standard 1D hydrostatic solar model atmospheres employed in the
analysis provide a realistic description of the line formation,
\item the adopted transition probabilities for the OH lines are correct,
\item the adopted microturbulence is appropriate for these partly
saturated OH A-X lines,
\item the continuum placement in the very crowded UV region of the OH lines
is correct,
\item local thermodynamic equilibrium (LTE) for the OH line formation and
equilibrium chemistry for the OH molecule formation are justified, and
\item LTE is good approximation for the Be\,{\sc ii} line formation.
\end{enumerate}
Addressing some of these lingering potential sources of errors is now possible.
Recently, the new generation of 3D hydrodynamical model atmospheres
(Stein \& Nordlund 1998; Asplund et al. 1999, 2000a;
Asplund \& Garc\'{\i}a P{\'e}rez 2001) has been applied to studies
of spectral line formation in the solar atmosphere for abundance determinations
(Asplund 2000; Asplund et al. 2000b, 2003a,b,c,d; Allende Prieto et al. 2001, 2002b;
Shchukina \& Trujillo Bueno 2001).
A particularly noteworthy finding from these studies is the substantial
downward revision of the solar photospheric O abundance to
log$\,\epsilon_{\rm O}=8.66\pm0.05$ (Asplund et al. 2003d). This new abundance has been
determined from a multitude of different spectral lines and for
the first time all O diagnostics yield consistent results:
the forbidden [O\,{\sc i}] line at 630\,nm, high-excitation,
permitted O\,{\sc i}, OH vibration-rotation lines and OH pure rotation lines.
In view of the significantly lower O abundance commonly accepted today
(Lodders 2003) compared
with the value believed to be correct at the time of the study by
Balachandran \& Bell (1998), it is clearly of importance to repeat their
analysis using the new, highly realistic, 3D hydrodynamical solar model atmosphere.
Such calculations are presented here, basically confirming the
findings of Balachandran \& Bell (1998) of significant ($\sim 50$\%) missing
UV-opacity and a photospheric Be abundance in agreement with
the meteoritic abundance.
Of the above-mentioned six uncertainties in the original
Balachandran \& Bell (1998) study, apparently now only the penultimate 
one remains which could possibly modify these conclusions.

\section{3D spectral line formation calculations}

The same 3D hydrodynamical solar model atmosphere which has previously
been confronted successfully with a range of observational diagnostics
(e.g. Asplund et al. 2000a) and used for abundance analysis purposes
(Asplund 2000; Asplund et al. 2000b, 2003a,b,c,d; Allende Prieto et al. 2001, 2002b;
Shchukina \& Trujillo Bueno 2001)
is employed here. For numerical details of the simulation, the reader
is referred to Stein \& Nordlund (1998), Asplund et al. (2000a) and
Asplund \& Garc\'{\i}a P{\'e}rez 2001).

The 3D spectral line formation calculations are performed under the
assumption of LTE for the ionization and excitation balances and for the
source function ($S_\nu = B_\nu$). Instantaneous chemical equilibrium
is furthermore assumed for the OH molecule formation.
The radiative transfer is solved for 17 different inclined directions
($N_\mu = N_\varphi = 4$ plus the vertical) using realistic
background continuous opacities (Gustafsson et al. 1975; Asplund et al. 1997)
and an equation-of-state which accounts for excitation, ionization
and molecule formation of the most important elements (Mihalas et al. 1988).
The photo-ionization cross-sections of Fe\,{\sc i} as calculated by
the Iron Project (Hummer et al. 1993) have not been included. It is likely that
the missing UV-opacity quantified below is indeed due to
Fe\,{\sc i} bound-free opacity (Bell et al. 2001).
The inclined rays are disk-integrated adopting a solar rotational velocity
of 1.8\,km\,s$^{-1}$ to produce flux profiles; the exact
choice of rotational broadening is however inconsequential for this study.
The instantaneous flux profiles are subsequently
averaged over a simulation time-sequence corresponding to about 50\,min
of solar time, which is sufficiently long to yield statistically significant
results.

\begin{table*}[t!]
\caption{The predicted line strengths in pm (=10\,m\AA ) for the OH A-X and Be lines
using 1D {\sc marcs} and Holweger-M\"uller solar model atmospheres calculated with
different combinations of microturbulence ($\xi_{\rm turb}$) and additional continuous
opacity over standard values ($\Delta \kappa_\nu^{\rm cont}/\kappa_\nu^{\rm cont}$).
The ``observed" equivalent widths used for the 3D calculations are those computed
using $\Delta \kappa_\nu^{\rm cont}/\kappa_\nu^{\rm cont} = 0.60$ (i.e. 60\% extra
opacity) and
$\xi_{\rm turb}=1.0$, as found by Balachandran \& Bell (1998).
\label{t:eqw1d}
}
\begin{tabular}{lcccccccc}
 \hline
line & $\chi_{\rm exc}$ & log\,$gf$ &
$\Delta \kappa_\nu^{\rm cont}/\kappa_\nu^{\rm cont}$&
\multicolumn{2}{c}{$W_\lambda$({\sc marcs})} &&
\multicolumn{2}{c}{$W_\lambda$(Holweger-M\"uller)}  \\
$ $[nm] & [eV] &&& \multicolumn{2}{c}{log\,$\epsilon_{\rm O} = 8.75$} &&
\multicolumn{2}{c}{log\,$\epsilon_{\rm O} = 8.91$}\\
        &      &&& \multicolumn{2}{c}{log\,$\epsilon_{\rm Be} = 1.40$} &&
\multicolumn{2}{c}{log\,$\epsilon_{\rm Be} = 1.40$}\\
\cline{5-6} \cline{8-9}\\
 & & &&
 $\xi_{\rm turb}=1.0$ &
 $\xi_{\rm turb}=1.2$ &&
 $\xi_{\rm turb}=1.0$ &
 $\xi_{\rm turb}=1.2$ \\
 & &&& [km\,s$^{-1}$] & [km\,s$^{-1}$]&& [km\,s$^{-1}$]& [km\,s$^{-1}$]
\smallskip\\
\hline\\
%\smallskip

OH 312.80 & 0.102 & $-2.425$ & 0.00 & 6.53 & 6.69 && 6.55 & 6.71 \\
          &       &          & 0.60 & 5.90 & 6.04 && 6.01 & 6.15 \\
OH 312.82 & 0.209 & $-2.074$ & 0.00 & 7.55 & 7.75 && 7.58 & 7.78 \\
          &       &          & 0.60 & 6.89 & 7.07 && 7.02 & 7.21 \\
OH 313.91 & 0.760 & $-1.563$ & 0.00 & 7.47 & 7.67 && 7.51 & 7.71 \\
          &       &          & 0.60 & 6.78 & 6.95 && 6.92 & 7.10 \\
OH 316.71 & 1.108 & $-1.544$ & 0.00 & 6.15 & 6.29 && 6.19 & 6.33 \\
          &       &          & 0.60 & 5.46 & 5.58 && 5.59 & 5.72 \\
Be\,{\sc ii} 313.10 & 0.000 & $-0.468$ & 0.00 & 8.79 & 8.91 && 8.63 & 8.74 \\
          &       &          & 0.60 & 7.63 & 7.73 && 7.63 & 7.73 \\
\hline
\end{tabular}
\end{table*}

A major advantage with the new generation of 3D hydrodynamical model atmospheres
is that the traditional free parameters of stellar spectroscopy (mixing length
parameters to describe convection and the micro- and macroturbulence for
line broadening) no longer are needed due to the self-consistently
calculated convective velocities and the corresponding Doppler shifts
(Asplund et al. 2000a,b; Asplund 2000; Allende Prieto et al. 2002a).
Thus, no microturbulence $\xi_{\rm turb}$
enters the 3D calculations presented herein,
although the corresponding 1D calculations have been performed with both
$\xi_{\rm turb} = 1.0$\,km\,s$^{-1}$ (the standard case as well as the
value adopted by Balachandran \& Bell 1998) and
$\xi_{\rm turb} = 1.2$\,km\,s$^{-1}$, to assess the uncertainties
attached to the adopted microturbulence in the original analysis of
Balachandran \& Bell (1998).

The molecular data for the OH and Be lines are well-known. Here a small
selection of four OH A-X lines are chosen (OH 312.80, 312.82, 313.91
and 316.71\,nm) covering the relevant range of excitation potential.
As will be evident later on, all lines suggest very similar multiplication
factors to the opacity at these wavelengths, and the exact choice of
OH lines is thus unimportant. The adopted $gf$-values for the OH A-X lines
are taken from Gillis et al. (2001), which are based on the most recent
theoretical calculations and laboratory measurements.
For completeness, it is noted that a dissociation energy of 4.392\,eV
(Sauval \& Tatum 1984) has been employed but it has no influence on the
conclusions obtained here.
The Be\,{\sc ii} 313.10\,nm
$gf$-value comes from the VALD database (Piskunov et al. 1995).
The relevant input data is listed in Table \ref{t:eqw1d}.

The study presented here is done strictly differentially between the
1D and 3D cases, thereby avoiding a direct comparison with observations.
Thus, in effect an intrinsic assumption for these calculations is that
the investigation of Balachandran \& Bell (1998) is done correctly
in terms of the comparison with observations within the framework of
1D model atmospheres and line formation.
The main reason to here avoid confrontation with observations is
that the current version of the 3D line formation code can only
treat lines of one element or molecule at a time, while the OH A-X
and Be lines are located in a very crowded spectral region with
contributions from many different species.
An advantage with the procedure adopted here is that the impact of the
3D model atmospheres and the new low solar O abundance can be
isolated without the possible confusion introduced by for example
continuum placement and choice of transition probabilities.
As there are all reasons to believe that the careful analysis
of Balachandran \& Bell (1998) was done correctly, we are still
in a position to evaluate the solar photospheric Be abundance
and the amount of missing UV-opacity.

As the predicted line shapes are intrinsically different for the 1D
and 3D cases even for the same line strengths, the here adopted
procedure is simplified to only use equivalent widths.
With the absence of direct observational confrontation, the observed
line strengths of the relevant OH A-X lines have been estimated
by redoing the analysis of Balachandran \& Bell (1998)
with the continuous opacities multiplied by a factor of 1.6.
The adopted solar O abundances for the calculations using
the 1D {\sc marcs} (Asplund et al. 1997) and the Holweger-M\"uller (1974)
solar model atmospheres are log$\,\epsilon_{\rm O}=8.75$ and
log$\,\epsilon_{\rm O}=8.91$, respectively, as found by
Balachandran \& Bell (1998) from the OH IR vibration-rotation lines with the
two model atmospheres.
In addition, the same calculations have been performed without
the additional continuous opacity as well as for the two values of the
microturbulence.
The predicted 1D equivalent widths of the OH A-X lines for the different
cases are listed in Table \ref{t:eqw1d} together with the corresponding
results for the Be\,{\sc ii} 313.10\,nm line.
The equivalent widths computed with a multiplication factor of 1.6 to
the continuous opacity will thus function here as substitutes for
the observed line strengths.

From a closer inspection of Table \ref{t:eqw1d}, it is clear that
the two 1D model atmospheres do not yield exactly the same ``observed"
equivalent widths, even if Balachandran \& Bell (1998)
claimed that their analysis yielded the same multiplication factor
to the continuous opacities in the two cases.
The reason for this relatively small discrepancy has not been identified but
can probably be traced to a combination of slightly different adopted
input data (continuous opacities etc),
model atmospheres (exact version of {\sc marcs} models, pressure-integrated
Holweger-M\"uller model atmosphere or not), overall solar chemical
composition, and numerical implementation of the 1D spectrum synthesis.
Assuming for the moment that the calculations with the Holweger-M\"uller
model atmospheres are correct with 60\% extra opacity, the same result
would be obtained for the {\sc marcs} case with about 50\% additional opacity
instead. It should be emphasized that this difference has no significant impact
on the conclusions regarding missing UV-opacity and the photospheric Be abundance.

\section{Results}

\begin{figure}[t]
\resizebox{\hsize}{!}{\includegraphics{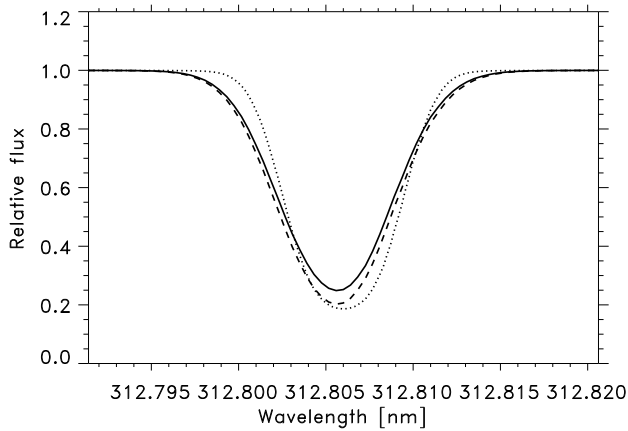}}
\caption{The dotted line denotes the OH 312.80\,nm line calculated with the
{\sc marcs} model atmosphere using a solar oxygen abundance of
log$\,\epsilon_{\rm O}=8.75$, a microturbulence of
$\xi_{\rm turb} = 1.0$\,km\,s$^{-1}$ and 60\% additional continuous opacity over
the standard value computed by the spectrum synthesis program. According to
Balachandran \& Bell (1998) this fits the observations very well, and is
here used as a substitute to the observed profile in this crowded
spectral region. The dashed line corresponds
to a 3D line calculation with log$\,\epsilon_{\rm O}=8.66$ and without any
extra opacity. The solid line represents the 3D results with the same
abundance but with 50\% extra continuous opacity, which yields
the same overall line strength as the here shown 1D profile. No micro- and macroturbulence
enter the 3D calculations but the Doppler shifts caused by the convective motions
make the line broader than the 1D case which is here shown without macroturbulence.
}
         \label{f:OH}
\end{figure}

Table \ref{t:eqw1d} and Table \ref{t:eqw3d} list the predicted line strengths
for the OH and Be\,{\sc ii} lines using the 1D and 3D model atmospheres, respectively.
As described above, the theoretical equivalent widths calculated with 1D model
atmospheres, a microturbulence of $\xi_{\rm turb} = 1.0$\,km\,s$^{-1}$ and
a multiplication factor of 1.6 to the continuous opacities around 313\,nm will
here function as the ``observed" values to be reproduced with the predictions
from the 3D spectrum synthesis. For the OH lines in 3D a solar O abundance of
log$\,\epsilon_{\rm O}=8.66$ has been adopted, as indicated from our recent
analysis of the OH vibration-rotation, the OH pure rotation,
the forbidden [O\,{\sc i}] lines at 630.0 and 636.3\,nm, and high-excitation,
permitted O\,{\sc i} lines (Asplund et al. 2003d).
It is then straightforward to compare the predicted equivalent widths using
the 3D hydrodynamical model atmosphere with different multiplication factors
to the continuous opacity with the ``observed" line strengths of the OH A-X lines.
It is immediately obvious that without additional continuous opacity over those
included in the spectrum synthesis package the OH lines are significantly too
strong (Fig. \ref{f:OH}). Using an average of the ``observed" values from the Holweger-M\"uller
and {\sc marcs} 1D model atmospheres, this missing opacity can be estimated
to be about 50\%, which is indeed very similar to the value of 60\% obtained by
Balachandran \& Bell (1998).

\begin{figure}[t]
\resizebox{\hsize}{!}{\includegraphics{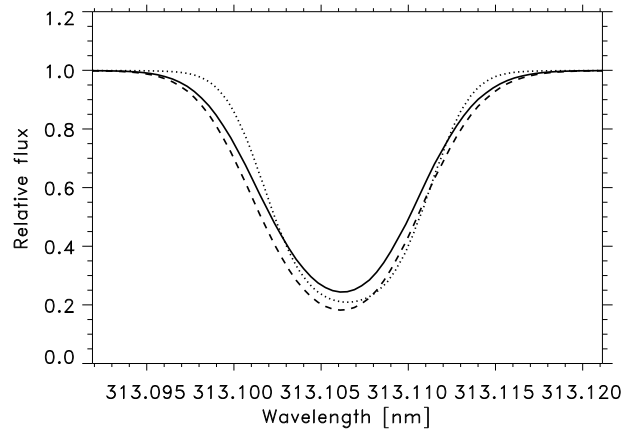}}
\caption{The dotted line denotes the Be\,{\sc i} 313.10\,nm line calculated with the
{\sc marcs} model atmosphere using a solar beryllium abundance of
log$\,\epsilon_{\rm Be}=1.40$, a microturbulence of
$\xi_{\rm turb} = 1.0$\,km\,s$^{-1}$ and 60\% additional continuous opacity, which
is required to fit the observations according to
Balachandran \& Bell (1998). The dashed line corresponds
to a 3D line calculation with log$\,\epsilon_{\rm Be}=1.38$ and without any
extra opacity. The solid line represents the 3D results with the same
abundance but with 50\% extra continuous opacity, which yields the same
line strength as the here shown 1D profile.
}
         \label{f:Be}
\end{figure}

The ``observed" equivalent width of the Be\,{\sc ii} 313.0\,nm line is
independent of whether the Holweger-M\"uller or the {\sc marcs} 1D model atmosphere
have been used to estimate it, as clear from Table \ref{t:eqw1d} (7.63\,pm=76.3\,m\AA ).
In 3D, without any additional continuous opacity this line strength would imply
a solar Be abundance of log$\,\epsilon_{\rm Be}=1.23$. However, as clear from
the comparison of the OH A-X lines, there are strong indications that the missing
UV-opacity amounts to about 50\%. Taking this extra opacity into account the
3D spectrum synthesis leads to accordingly weaker lines, yielding a
higher solar Be abundance of log$\,\epsilon_{\rm Be}=1.38$ (Fig. \ref{f:Be}).
Combined with an estimate of the errors in the missing opacity and the overall
agreement between predicted and observed line profiles (Balachandran \& Bell 1998),
the final result is thus log$\,\epsilon_{\rm Be}=1.38 \pm 0.09$.
This best estimate using a 3D model atmosphere is indistinguishable from the
meteoritic value of log$\,\epsilon_{\rm Be}=1.41 \pm 0.08$ (Lodders 2003)
given the uncertainties in both the photospheric and meteoritic analyses.
This is even more true in the light of the proposed downward revision by 0.04\,dex
to the meteoritic abundance scale caused by the re-analysis of the photospheric
Si abundance on which the absolute meteoritic abundance scale is anchored
(Asplund 2000), which would bring the meteoritic value down to
log$\,\epsilon_{\rm Be}=1.37 \pm 0.08$.

The findings of Balachandran \& Bell (1998) of significant missing UV-opacity
and no Be depletion within the solar convection zone throughout the 4.6\,Gyr
of solar evolution until today is therefore confirmed by this analysis
based on a realistic 3D model atmosphere rather than 1D models.

\begin{table}[t!]
\caption{The predicted line strengths in pm for the OH A-X and Be lines
using a 3D hydrodynamical solar model atmosphere. The OH lines are
computed for log$\,\epsilon_{\rm O}=8.66$ and the Be line for
log$\,\epsilon_{\rm Be}=1.38$.
\label{t:eqw3d}
}
\begin{tabular}{lcccc}
 \hline
line & $\chi_{\rm exc}$ & log\,$gf$ &
$W_\lambda$ & $W_\lambda$ \\
$ $ [nm] & [eV] & & 
$\Delta \kappa_\nu^{\rm cont}/\kappa_\nu^{\rm cont}$ &
$\Delta \kappa_\nu^{\rm cont}/\kappa_\nu^{\rm cont}$ \\
& & & $=0.00$ & $=0.50$\\
 \hline
OH 312.80 & 0.102 & $-2.425$ & 6.38 & 5.90 \\
OH 312.82 & 0.209 & $-2.074$ & 7.42 & 6.91 \\
OH 313.91 & 0.760 & $-1.563$ & 7.29 & 6.75 \\
OH 316.71 & 1.108 & $-1.544$ & 5.87 & 5.34 \\
Be\,{\sc ii} 313.10 & 0.000 & $-0.468$ & 8.62 & 7.65 \\
\hline
\end{tabular}
\end{table}

\section{Remaining uncertainties}

In this section, uncertainties which may have a bearing on the
findings are discussed.

{\bf Model atmospheres:}
Available evidence from various confrontations between the predictions from
the here employed 3D solar simulation with observed diagnostics all
suggest that indeed the new 3D hydrodynamical model atmospheres
are highly realistic, representing a significant improvement over existing
1D models. It is reassuring that the main conclusions regarding the missing
UV-opacity and the solar Be abundance are robust against choice of
model atmosphere, whether in 1D or in 3D.

{\bf Line broadening:}
It should be noted that the line strengths of the OH A-X lines are somewhat
sensitive to the adopted microturbulence in 1D. Had instead
Balachandran \& Bell (1998) employed a larger microturbulence, they would
have obtained slightly smaller multiplication factors to the continuous opacity,
as clear from Table \ref{t:eqw1d}. Due to the differential nature of this
study relative to that of Balachandran \& Bell (1998), the multiplication factor
estimated here using a 3D solar model atmosphere would then have been
adjusted slightly accordingly. For reasonable values of the microturbulence, however,
the changes are small to the estimated amount of missing UV-opacity and the
derived solar Be abundance. Available solar analyses all indicate that a
microturbulence of $1.0 \pm 0.1$\,km\,s$^{-1}$ for flux spectra
is appropriate when relying on
1D model atmospheres. The uncertainties attached to the exact choice of
microturbulence can therefore in this case be safely ignored.

{\bf Transition probabilities:}
Conceivably, the adoption of erroneous transition probabilities for either
the OH vibration-rotation or the OH A-X lines could be misleading,
yielding incorrect estimates
of the amount of missing UV-opacity, if any, and consequently the derived
solar Be abundance. Fortunately, the necessary $gf$-values for both sets
of lines are accurately known from theoretical calculations and
laboratory measurements (Goldman et al. 1998; Gillis et al. 2001),
which are very similar to those
used by Balachandran \& Bell (1998) in their study.
The uncertainties in the transition probabilities are sufficiently small
not to compromise the conclusions presented here.
Similarly, as long as consistent partition function and molecular equilibrium
constants are employed, the results are independent of such input data.

{\bf Continuum placement:}
A potentially more serious shortcoming of the study of Balachandran \& Bell (1998)
and thus also the present analysis, is the exact placement of the continuum level.
If the observed continuum is estimated to be too low,
a too low abundance is derived, in particular given
the fact the relevant lines for these studies are at least partly saturated.
The region around the OH A-X and Be lines at 313\,nm is notoriously crowded and
it is difficult to identify wavelength windows which traces the continuum.
The problem is compounded by the fact that many of the
lines in this region remain unidentified, rendering even the use of
spectrum synthesis to locate small continuum portions uncertain.
As this study does not involve direct comparison with observations, it relies
entirely on that this part in the original Balachandran \& Bell (1998) work was
carried out properly. From the relatively small wavelength regions shown in
their figures it appears however that the continuum placement is essentially
correct, which should minimize the uncertainties for the final results.

{\bf Non-LTE line formation for OH lines:}
While LTE is in all likelihood an excellent approximation for the formation of
the OH vibration-rotation lines, this is not guaranteed for the OH A-X lines
(Hinkle \& Lambert 1975). To date, very little work on non-LTE line formation
has been invested for molecules with the notable exception of CO lines
(e.g. Uitenbroek 2000). For OH A-X lines, the author is only aware of the
very preliminary study of Asplund \& Garc\'{\i}a P{\'e}rez (2001) based on
only a two-level OH model. A fundamental obstacle in this respect is the
lack of data for excitation and ionization by collisions with electrons and hydrogen
atoms, even more so than for the corresponding case of atoms.
With their very simplified approach Asplund \& Garc\'{\i}a P{\'e}rez (2001) indeed
found the possibility of significant departures from LTE for the OH A-X lines,
amounting to about 0.2\,dex in terms of oxygen abundance for the solar case in 1D.
If this significant weakening of the predicted OH A-X lines in non-LTE is correct,
the conclusions regarding missing UV-opacity must be revised. Indeed,
a non-LTE effect of 0.2\,dex more than compensates for the effect of extra
opacity. For example with an O abundance of log$\,\epsilon_{\rm O}=8.55$, the
predicted equivalent width of the OH 312.80\,nm line with the 1D {\sc marcs} model
atmosphere, $\xi_{\rm turb} = 1.0$\,km\,s$^{-1}$ and no extra opacity is
5.60\,pm, i.e. weaker than the ``observed" line strength. In other words, this would
at face value imply that there is no missing UV-opacity but rather that the
known opacity sources predict {\em too much} opacity in this particular wavelength
region. However, the reader is urged to be cautious and not over-interpreting
the very preliminary non-LTE results of Asplund \& Garc\'{\i}a P{\'e}rez (2001).
It is likely that the inclusion of more
levels in the OH model would more efficiently thermalize the molecule,
bringing the results closer to LTE.
It is clear that a more careful non-LTE investigation of the OH line formation
is necessary to resolve this issue.

{\bf Non-equilibrium chemistry for OH molecule formation:}
Another crucial assumption for these types of studies is the use of
instantaneous equilibrium chemistry for the molecule formation.
In reality the time-scale for molecule formation is finite and
if this approaches the typical convective time-scale for up- and down-flows
in the solar atmospheres the number density of molecules will in
general be lower than expected from equilibrium chemistry.
This has been shown by Asensio Ramos et al. (2003) 
for the case of the CO molecule in hydrodynamical models of the solar chromosphere.
As for non-LTE line formation for molecules, little work has been done
in this potentially crucial area. The available evidence, however, indicates that
equilibrium chemistry is valid for OH, at least in the atmospheric
layers corresponding to the formation heights of the OH vibration-rotation
and A-X lines in the Sun (S\'anchez Almeida et al. 2001; Asensio Ramos et al. 2003;
Asensio Ramos \& Trujillo Bueno 2003).
In addition, these effects only influence the present study in as far as
the two types of lines are formed in different layers with a differing
degree of appropriateness for the assumption of equilibrium chemistry:
if both the vibration-rotation and A-X lines have the same typical
formation heights any non-equilibrium molecule formation will affect
both similarly, leaving the estimate of the missing UV-opacity
and the solar Be abundance intact.

{\bf Non-LTE line formation for Be lines:}
Finally, possible departures from LTE for the line formation of the
Be\,{\sc ii} lines can of course influence the derived solar Be abundance.
Fortunately, the two main non-LTE effects for these lines -- over-ionization
and over-excitation -- essentially compensate each other for the solar case,
coincidentally leaving the non-LTE Be abundance very similar to the LTE value:
$\Delta($log$\,\epsilon_{\rm Be}) < 0.02$\,dex
(Garc\'{\i}a P{\'e}rez et al., in preparation).
Thus, the finding of no significant Be depletion is safe to departures from
LTE in the Be line formation.

\section{Conclusions}

In the present article a re-analysis of the work by Balachandran \& Bell (1998)
has been performed employing a state-of-the-art 3D hydrodynamical solar
model atmosphere (Asplund et al. 2000a) instead of traditional 1D hydrostatic
models. Several of the uncertainties attached to the analysis by
Balachandran \& Bell (1998) are thus removed and the important results placed on a
more firm footing. In addition, several of the outstanding sources of
uncertainties have here been discussed at some length and, with one
notable exception, found to be in most likelihood unimportant for the basic findings.
The conclusions by Balachandran \& Bell (1998) of a significant missing UV-opacity 
around 313\,nm and a photospheric Be abundance in
very close agreement with the meteoritic value measured in CI-chondrites are
thereforeconfirmed by this study based on a more
realistic model atmosphere. According to this study, the
additional opacity amounts to about 50\% in this particular wavelength region,
which is only slightly smaller than Balachandran \& Bell (1998) estimated
using 1D model atmospheres (about 60\%).

Although this study does not shed further light on the nature of the
opacity source causing the missing UV-opacity it is likely that it
can be traced to photo-ionization of Fe\,{\sc i}. Using data from
the Iron Project, Bell et al. (2001) indeed found that such
unaccounted for Fe\,{\sc i} bound-free cross-sections are close to explaining the
missing 60\% continuous opacity estimated by Balachandran \& Bell (1998).
This will ring even more true with the slightly smaller multiplication
factors derived here.

The situation related to the depletion of the light elements in the solar
photosphere can thus be summarized as follows: while lithium has been
destroyed by a factor of about 140 (Lodders 2003; Asplund et al. 2003a),
neither the photospheric beryllium nor the boron
abundances appear to have been significantly modified throughout the solar
evolution to the present day.
As described in the Introduction, no standard model of stellar evolution
can account for this depletion pattern. Furthermore, is poses a severe challenge
for non-standard models such as slow rotational-induced mixing below the
convection zone. Indeed, it is a benchmark which any such models must
successfully be tested against before they can be trusted for other types of stars.

\begin{acknowledgements}
This work has been supported
by the Swedish Research Council (F990/1999) and
the Australian Research Council (DP0342613).
The author gratefully acknowledges financial support from
the local organizing committee of the conference {\em CNO in
the Universe} where these results were presented.

\end{acknowledgements}

\end{document}